\documentclass[aps,prd,superscriptaddress,showpacs,altaffilletter,twocolumn]{revtex4}
\usepackage[dvips]{graphicx}
\def\to{\rightarrow}

\def\AJL{{Ap. J. Lett.} }

\def\FP{{Fortschr. Physik} }
\def\GRG{{Gen. Rel. Grav.} }

\def\JHEP{{JHEP} }

\def\MNRAS{{Mon. Not. R. Ast. Soc.} }

\def\NC{{Il Nuovo Cimento} }
\def\NP{{Nucl. Phys.} }
\def\PL{{Phys. Lett.} }
\def\PR{{Phys. Rev.} }
\def\PRL{{Phys. Rev. Lett.} }

\def\be{\beta}

\def\lsim{\mathrel{\rlap{\lower4pt\hbox{\hskip1pt$\sim$}}
    \raise1pt\hbox{$<$}}}
\def\gsim{\mathrel{\rlap{\lower4pt\hbox{\hskip1pt$\sim$}}
    \raise1pt\hbox{$>$}}}
\def\sqr#1#2{{\vcenter{\vbox{\hrule height.#2pt
         \hbox{\vrule width.#2pt height#1pt \kern#1pt
         \vrule width.#2pt}
         \hrule height.#2pt}}}}

\def\be{\begin{equation}}
\def\ee{\end{equation}}
\def\bea{\begin{eqnarray}} 
\def\eea{\end{eqnarray}}

\begin{document}
\author{Luis P. Chimento}
\email{chimento@df.uba.ar}
\affiliation{Dpto. de F\'\i sica, Facultad de Ciencias Exactas y Naturales,  Universidad de Buenos Aires, Ciudad Universitaria,  Pabell\'on I, 1428 Buenos Aires, Argentina}
\author{Ruth Lazkoz}
\email{wtplasar@lg.ehu.es}
\affiliation{Fisika Teorikoa, Zientzia eta Teknologia Fakultatea, Euskal Herriko Unibertsitatea, 644 Posta Kutxatila, 48080 Bilbao}
\title{Large-scale inhomogeneities in modified Chaplygin gas cosmologies}

\begin{abstract} 
We extend the homogeneous modified Chaplygin cosmologies to large-scale perturbations by formulating a Zeldovich-like approximation. We show that the model interpolates between  an epoch with a soft equation of state and a de Sitter phase, and that in the intermediate regime its matter content is simply the sum of dust and a cosmological constant. We then study how the large-scale inhomogeneities evolve and compare the results with cold dark matter  (CDM), $\Lambda$CDM and generalized Chaplygin scenarios. Interestingly, we find that  like the latter, our  models  resemble $\Lambda$CDM.
\end{abstract}

\pacs{98.80.Cq}
\maketitle
\section{Introduction}
According to increasing astrophysical indicia,
 the evolution of the Universe seems to be largely governed by dark energy  with negative
 pressure together with pressureless cold dark matter (see \cite{Sahni:2004ai} for the latest review) in a two to one proportion. However, little is know about the origin of either component, which in the standard cosmological model would
 play very different roles: dark matter would be responsible for matter clustering, whereas dark energy \cite{Peebles:2002gy} would account for accelerated expansion. Several candidates
 for dark energy haven proposed and confronted with observations: a purely cosmological constant, quintessence with a single field (see \cite{quintessence} for earliest papers) or two coupled fields \cite{quint_two},  k-essence scalar fields, and phantom energy \cite{phantom}. Interestingly, a  bolder alternative presented recently suggests that
an effective dark energy-like equation of state could be due to averaged quantum effects
 \cite{onemli}.

The lack of information regarding the provenance of dark matter and dark energy
allows for speculation with the economical and aesthetic idea that a single component acted in fact as both dark matter and dark energy. 
The unification of those two components has risen a considerable theoretical interest, because on the one hand  model building becomes  considerable simpler, and on the other hand such unification implies the existence of an era during which the energy densities of dark matter and dark energy are  strikingly similar. 

One possible way to achieve that unification is through a particular k-essence fluid, the generalized Chaplygin gas \cite{Bentogen}, with the exotic equation of state 
\begin{equation}
p=-\frac{A}{\rho^{\beta}}\end{equation} where constants $\beta$ and $A$ satisfy respectively $0<\beta\le 1$ and $A>0$.
 Using the energy conservation equation and the Einstein equation $3H^2=\rho$ one obtains  the evolution 
 \begin{equation} 
 3H^2=\left(A+\frac{B}{a^{3(1+\beta)}}\right)^{1/(1+\beta)},
 \end{equation}
 where as usual $a$ is the scale factor, $H=\dot a/a$ and $B>0$ is an integration constant. This  model  interpolates between  a $\rho\propto a^{-3 }$ evolution law at early times and $\rho\simeq\rm{cons.}$ at late times (i.e. the model is dominated by dust  in its early stages and by vacuum energy in its late  history). In the  intermediate regime the matter content of the model can be approximated by
 the sum of a cosmological constant an a fluid with a soft equation
 of state $p=\beta \rho$. The traditional Chaplygin gas \cite{Kamenshchik}
 corresponds to $\beta=1$ (stiff equation of state).
 
Another possibility which has emerged recently is the modified Chaplygin gas (MCG) \cite{chimento}. It is characterized by
\begin{eqnarray}
&&\rho=  \left(A + {B \over a^{3}}\right)^{\alpha/(\alpha-1)}.
\label{eqstate}\\
&&p= \frac{1}{ \alpha-1}\left(\rho-\alpha A \rho^{1/\alpha}\right),
\label{effecden}
\end{eqnarray}
with $\alpha>0$ a constant. 

Alternatively, such evolution can be  seen as coming from a modified gravity approach, along the lines of the Dvali-Gabadadze-Porrati \cite{dgp}, Cardassian \cite{cardas} and Dvali-Turner \cite{dt} models.  In those works the present acceleration of the universe is not attributed to an exotic component in the Universe, but to modifications in gravitational physics at subhorizon scales. Following the proposal by \cite{barsen}, an evolution like  that arising from (\ref{eqstate})
in standard gravity, could alternatively be obtained  in the modified gravity picture for a pure dust dust configuration under the modification   
\begin{equation}
3H^2=(A+\rho_m)^{\alpha/(\alpha-1)},
\end{equation}
where $\rho_m\propto a^{3}$.

Modified Chaplygin cosmologies  with $\alpha> 1$ are transient models which interpolate between a $\rho\propto a^{-3 \alpha/(\alpha-1)}$ evolution law at early times  and a de Sitter phase at late times, but interestingly the
matter content at the intermediate stage is a mixture
of dust and a cosmological constant. The sound speed
for the modified
Chaplygin gas \cite{chimento} becomes
\be
\label{cs}
c_s^2=\frac{1}{\alpha-1}\left[1-A\rho^{(1-\alpha)/\alpha}\right].
\ee

The observational 
 tests of traditional and generalized Chaplygin models are numerous.  
Several teams have analyzed the compatibility of those models with the Cosmic Microwave Background Radiation (CMBR) peak location and amplitude \cite{cmbobs}, supernovae data 
\cite{supobs} and gravitational lensing statistics \cite{lensobs} . The main results of those papers can be summarized as follows: models with $\beta>1$ 
and some small curvature (positive or negative) are favored over the $\Lambda$CDM model, and Chaplygin cosmologies are much likelier  as dark energy models than as unified dark matter models.

In what regards the modified Chaplygin gas, it has only been tested observationally in \cite{oscillations}, using the most updated and reliable compilation of supernovae data so far: the Gold dataset by Riess et al. \cite{riess}. By means of a statistical test which depends no only on $\chi^2_{min}$ (as in usual procedures) bu also on the number of parameters of the parametrization of the Hubble factor as a function of redshift, it was concluded that the   modified Chaplygin gas cosmologies give better fits than   usual and generalized  Chaplygin cosmologies.

In this paper we shall be concerned with the evolution  of large-scale inhomogeneities in modified Chaplygin cosmologies. This is an  issue of interest because  candidates for the dark matter  and dark matter unification will only be valid if they ensure that initial perturbations can evolve into a deeply nonlinear
regime to form a gravitational condensate of super-particles that can act like cold
dark matter. 
Here we will follow the covariant and sufficiently general Zeldovich-like non-perturbative
approach  given in \cite{Bilic}, because it can be adapted to any balometric or parametric equation of state. Our results indicate that our model fits well in the standard structure formation scenarios, and  we find, in general, a fairly similar behavior to generalized Chaplygin models \cite{Bentogen}

\section{The Model} 

For the modified Chaplygin gas described by Eqs. 
(\ref{eqstate}) and (\ref{effecden})the effective equation of state in the intermediate regime
between the dust dominated phase and the de Sitter phase can be obtained
expanding Eqs. (\ref{effecden}) and (\ref{eqstate}) in powers of $Ba^{-3}$,
we get

\begin{eqnarray}
&&\rho= A^{\alpha/(\alpha-1)} + A^{1/(\alpha-1)}\frac{\alpha\,B} {(\alpha-1)a^3  } +{\cal O}\left(\frac{B^2}{a^{6}}\right)\label{rhoapprox}\\
&& p= -A^{\alpha/(\alpha -1)}+{\cal O}\left(\frac{B^2}{a^{6}}\right)\label{papprox},
\end{eqnarray}
which corresponds to a mixture of vacuum energy density
$A^{\alpha/\alpha-1)}$,
presureless dust and other perfect fluids which dominate at the very beginning
of the universe. In the intermediate regime the modified Chaplygin gas behaves
as dust at the time where the energy density satisfies the condition
$\rho=A^{\alpha/(\alpha-1)}$. At very early times the equation of
state parameter $w\equiv p/\rho$ becomes
\begin{equation}
w\simeq c_s^2\simeq\frac{1}{\alpha-1},\label{quasidust}
\end{equation}
so that for very large $\alpha$ the dust-like behavior is recovered.

The next step is to investigate what sort of cosmological model arises when we
consider a slight inhomogeneous modified Chaplygin cosmologies. For a general
metric $g_{\mu \nu}$, the proper time $d\tau = \sqrt{g_{00}} dx^0$, and
$\gamma \equiv -g/g_{00}$ as the determinant of the induced $3$-metric, one
has

\be
\gamma_{i j} = {g_{i0}g_{j0} \over g_{00}} - g_{ij}~.
\label{inducedmetric}
\ee
In the first approximation it will be interesting to investigate the
contribution of inhomogeneities introduced in the modified Chaplygin gas
through the expression
\be
\rho =  \left(A + \frac{B}{\sqrt{\gamma}}\right)^
{\alpha/(\alpha-1)}~.
\label{genevoldeninh}
\ee  The latter result suggests that the evolution of inhomogeneities
can be studied using the Zeldovich method through the deformation tensor 
\cite{Bilic,Matarrese,Peebles}:
\be
D_{i}^{j} = a(t) \left(\delta_{i}^{j} - b(t) 
{\partial^2 \varphi(\vec{q}) \over \partial q^i \partial q^j}\right)~,
\label{defromationt}
\ee
where $b(t)$ parametrizes the time evolution of the inhomogeneities and $\vec{q}$ are generalized Lagrangian coordinates so that 
\be
\gamma_{i j} = \delta_{mn}D_{i}^{m}D_{j}^{n}~,
\label{inducedmetric1}
\ee     
and $h$ is a perturbation
\be
h = 2 b(t) {\varphi_{,i}}^i~.
\label{hevol} 
\ee
Hence, using the equations above and Eqs. (\ref{rhoapprox}) and
(\ref{papprox}),  it follows that
\begin{eqnarray}
&&\rho \simeq \bar \rho (1 + \delta)~,
\\
&&p \simeq \frac{1}{\alpha-1}\left({{\bar\rho }} - A\,\alpha \,{{{\bar\rho \,}}}^{1/\alpha} + 
  \delta \,\left( {{\bar\rho }} - A\,{{{\bar\rho \,}}}^{1/\alpha} \right)\right)~,
\label{pert}
\end{eqnarray}
where $\bar \rho$ is given by Eq. (\ref{eqstate}) and the density contrast
$\delta$ is related to $h$ through 
\be
\delta =  {h \over 2} (1 + w)~,
\label{delta}
\ee 
where $w\equiv\bar p/\bar \rho$. Finally, after some algebra we get
\be
\bar p=\bar \rho\left(w + \frac{\left( 1 + w \right) \delta }{\alpha }\right).
\label{omega}
\ee 

Now, the metric (\ref{inducedmetric1}) leads 
to the following $00$ component of the Einstein equations:
\be
- 3 {\ddot{a} \over a} + {1 \over 2} \ddot{h} + H \dot{h} = 
4 \pi G \bar \rho \left(1 + 3w + \left( 1 + \frac{3\left( 1 + w \right) }{\alpha } \right) \delta\right)~,
\label{Einstein}
\ee 
where the unperturbed part of this equation corresponds 
to the Raychaudhuri equation
\be
- 3 {\ddot{a} \over a} = 4 \pi G \bar \rho (1 + 3 w)~.
\label{Raychaudhuri}
\ee 

Using the Friedmann equation for a flat spacetime
$
H^2= {8 \pi G} \bar \rho/3$, one can rewrite
Eq. (\ref{Einstein})   as a differential equation for $b(a)$:

\be
{2 \over 3} a^2 b'' + (1 - w) a b' - (1 + w) \left( 1 + \frac{3\left( 1 + w \right) }{\alpha } \right) b = 0~,
\label{bprimes}
\ee
where the primes denote derivatives with respect to the scale-factor, $a$.



 An expression for $w$ as a function of the scale-factor can be derived from
 Eqs. (\ref{eqstate}) and  (\ref{effecden}):
\begin{equation}w(a)= \frac{B - (\alpha-1)A{a}^3 }
  {\left(\alpha -1 \right) \left( B + A{a}^3 \right) }.
  \end{equation}
  The latter must be conveniently recast in terms of the fractional vacuum
and matter  energy densities. This can be done by using 
\begin{equation}\lim_{\alpha\to\infty}\rho=A+\frac{B}{a^3}\label{limit}\end{equation}
combined with
\begin{equation}
H^2=H_0^2\left(\Omega_{m0}\left(\frac{a_0}{a}\right)^3+{\Omega_{\Lambda 0}}
\right).
\end{equation}
where $H_0$ and $a_0$ are, respectively, the current value of the Hubble
and scale factor, and $\Omega_{\Lambda 0}$ and $\Omega_{m0} $ are, respectively, the fractional vacuum
and matter  energy densities today. Setting $a_0=1$ we  obtain
\begin{equation}w(a)= \frac{\Omega_{m0} - (\alpha-1)\Omega_{\Lambda 0}{a}^3 }
  {\left(\alpha -1 \right) \left( \Omega_{m0} + \Omega_{\Lambda 0}{a}^3 \right) },
  \label{wparam}
  \end{equation}
  and consistently
  \begin{equation}\lim_{\alpha\to\infty}w(a)= -\frac{ \Omega_{\Lambda 0}{a}^3 }
  { \Omega_{m0} + \Omega_{\Lambda 0}{a}^3  }.
  \end{equation}

\begin{figure}[t]
\begin{center}
\hspace{0.4cm}\includegraphics[width=7cm,height=4.5cm]{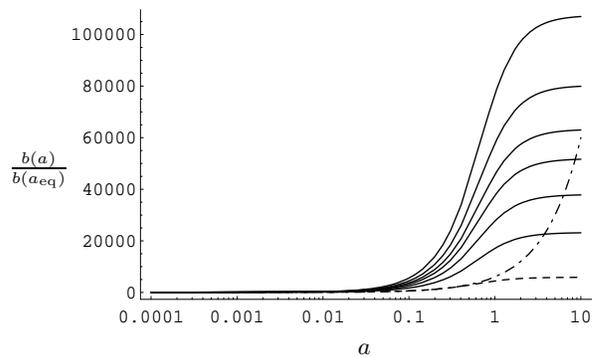}
\put(-222,62){$\frac{b(a)}{b(a_{\rm{eq}})}$}
\put(-90,-6){$a$}
\end{center}
\caption{Evolution of ${b(a)}/{b(a_{\rm{eq}})}$ for the modified Chaplygin gas for 
$\alpha=10,11,12,13,15,20$ (continuous lines) as compared with  $\Lambda$CDM (dashed line)
and CDM (dashed-dotted line). Lower curves correspond to higher values of $\alpha$.}
\label{evbsmall}\end{figure}

\begin{figure}[h]
\begin{center}
\hspace{0.4cm}\includegraphics[width=7cm,height=4.5cm]{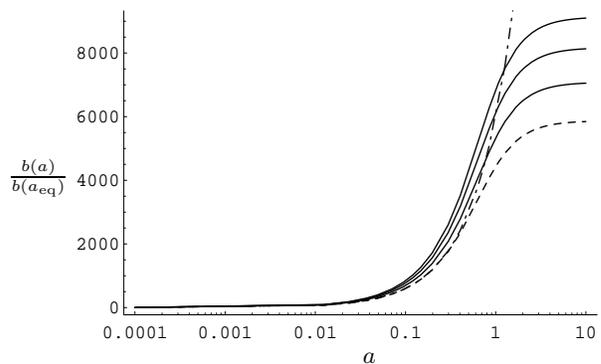}
\put(-224,62){$\frac{b(a)}{b(a_{\rm{eq}})}$}
\put(-90,-6){$a$}
\end{center}
\caption{Evolution of ${b(a)}/{b(a_{\rm{eq}})}$ for the modified Chaplygin gas for 
$\alpha=60,80,140$ (continuous lines) as compared with  $\Lambda$CDM (dashed line)
and CDM (dashed-dotted line). Lower curves correspond to higher values of $\alpha$.}
\label{evblarge}\end{figure}

We have used this expression to  integrate Eq. (\ref{bprimes}) numerically,
for different values of $\alpha$, and taking $\Omega_m=0.27$ and $\Omega_{\Lambda}=0.73$
\cite{bennett}.  We have set $a_{eq} = 10^{-4}$ 
for matter-radiation equilibrium (while keeping $a_{0} = 1$  at present), and our initial condition is $b'(a_{eq}) = 0$. Our results are shown in figures 
\ref{evbsmall} and \ref{evblarge}.

We find that 
modified Chaplygin scenarios start differing from the $\Lambda$CDM  
only recently ($z \simeq 1$) and that, in any case, they yield a 
density contrast that closely 
resembles, for any value of $\alpha > 1$, the standard CDM 
before the present. 
Notice that $\Lambda$CDM corresponds effectively 
to using Eq. (\ref{limit}) 
and removing the  
factor $(1 + 3(1+w)/ \alpha)$ in Eq. (\ref{bprimes}). Figures \ref{evbsmall} and \ref{evblarge} 
show also that, for any value of $\alpha$, $b(a)$ saturates 
as in the  $\Lambda$CDM case.

\begin{figure}[t!]
\begin{center}
\hspace{0.45cm}\includegraphics[width=7cm,height=4.5cm]{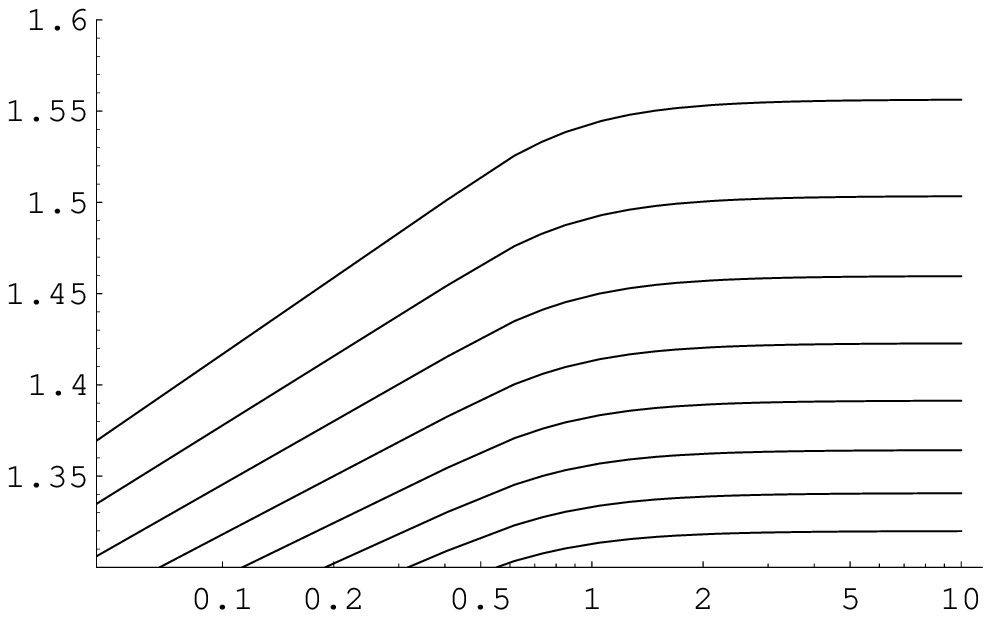}
\put(-228,62){$\frac{\delta_{\rm{mChap}}}{\delta_{\Lambda\rm{CDM}}}$}
\put(-90,-6){$a$}
\end{center}
\caption{Evolution of ${\delta_{\rm{mChap}}}/{\delta_{\Lambda\rm{CDM}}}$ for the modified Chaplygin gas for
$\alpha=60,65,70,75,80,85,90,95$ (continuous lines) as compared with  $\Lambda$CDM (dashed line). Lower curves correspond to higher values of $\alpha$.}
\label{ratiofig}\end{figure}
In what regards the density contrast, $\delta$, 
using Eqs. (\ref{hevol}), (\ref{delta}) and (\ref{wparam}) one can deduce 
that the ratio between this quantity in the modified Chaplygin and the $\Lambda$CDM 
scenarios is simply given by
\be
{\delta_{\rm{mChap}} \over \delta_{\Lambda\rm{CDM}}} = {b_{\rm{mChap}} \over
b_{\Lambda \rm{CDM}}} \frac{\alpha}{\alpha-1}~,
\label{ratio}
\ee 
and its behavior is depicted in figure
\ref{ratiofig}. We find that it asymptotically evolves to a constant value.

\begin{figure}[t!]
\begin{center}
\hspace{0.5cm}\includegraphics[width=7cm,height=4.5cm]{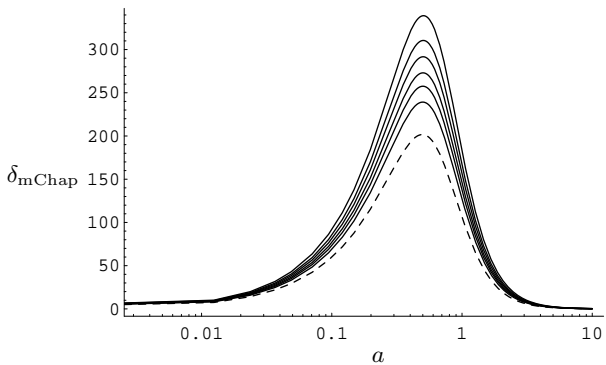}
\put(-228,62){${\delta_{\rm{mChap}}}$}
\put(-90,-6){$a$}
\end{center}
\caption{Evolution of ${\delta_{\rm{mChap}}}$ for the modified Chaplygin gas for
$\alpha=50,60,70,85,105,150$ (continuous lines) as compared with  $\Lambda$CDM (dashed line). Lower curves correspond to higher values of $\alpha$.}
\label{contrast}\end{figure}

  Now, in figure 
\ref{contrast}, we have plotted $\delta$  as a function of $a$ for different 
values of $\alpha$. As happens in the
 the traditional \cite{Bilic,Fabris} and generalized Chaplygin models, 
in our models the density contrast decays for large $a$ also.
\section{Discussion and conclusions}

Using a Zeldovich-like approximation,  we have studied the evolution of large-scale perturbations in a recently proposed theoretical framework for the unification of dark matter and dark energy: the so-called modified 
Chaplygin cosmologies \cite{chimento}, with equation of state $$p= \frac{1}{ \alpha-1}\left(\rho-\alpha A \rho^{1/\alpha}\right),$$ with 
$ \alpha > 1$. This model  evolves 
from a phase that is initially dominated
by non-relativistic matter to a phase that is 
asymptotically de Sitter.  The intermediate 
regime  corresponds to a phase where the effective equation 
of state is 
given by $p =0$ plus a cosmological constant. We have estimated the fate of the inhomogeneities 
admitted in the model and shown that these evolve consistently 
with the observations as the density contrast they introduce is 
smaller than the one typical of CDM scenarios. 

On general grounds, the pattern of evolution of perturbations follows is similar to the one in the $\Lambda$CDM models and in generalized Chaplygin cosmologies, and therefore our represent plausible alternatives alternatives

As usual, in modified Chaplygin cosmologies, the equation of state parameter $w$  can be expressed in terms of the scale factor and a free parameter $\alpha$, and the value of the latter can be chosen so that the model resembles as much as desired the $\Lambda$CDM model. 

It would be very interesting to deepen in the  comparison between modified and generalized Chaplygin models, particularly from the observational point of view  (as already done in \cite{oscillations}). 
It would also be worth generalizing our study by going beyond the Zeldovich approximation, to   incorporate the effects of finite sound speed. This can be done by generalizing the spherical model to incorporate the Jeans length as in 
\cite{lindebaum}. Alternatively, following \cite{decomposition} one could investigate whether the modified Chaplygin
admits an unique decomposition into dark energy and dark matter, and if that were the case
then  study structure 
formation and show that difficulties associated to unphysical oscillations or blow-up in the matter
power spectrum can be circumvented.
We hope this will be addressed in future works.

\begin{acknowledgments}

L.P.C. is partially funded by the University of Buenos Aires  under
project X224, and the Consejo Nacional de Investigaciones Cient\'{\i}ficas y
T\'ecnicas under proyect 02205.  R.L. is supported by  the University of the Basque Country through research grant 
UPV00172.310-14456/2002 and by the Spanish Ministry of Education and Culture  through research grant  FIS2004-01626. 
\noindent

\end{acknowledgments}

\end{document}